\documentclass[aps,prl,superscriptaddress,showpacs,preprintnumbers,amsmath,amssymb]{revtex4}
\usepackage[dvipdfmx]{graphicx}
\usepackage{dcolumn}
\usepackage{subfigure}
\usepackage{multirow}
\usepackage{comment}
\usepackage{color}

\begin{document}

\preprint{\vbox{ \hbox{   }
                 \hbox{BELLE-CONF-1501}
               % \hbox{ICHEP2008-xx}
               % \hbox{hep-ex nnnn, if available}
}}

\title{ \quad\\[1.0cm] 
Measurement of the amplitude ratio of $B^0 \to D^0K^{*0}$ and $B^0 \to \bar{D^0}K^{*0}$ decays with a model-independent Dalitz plot analysis using $D\to K_S^0\pi^+\pi^-$ decays}

%%% Paper:
%%% Journal:  2015 Conference Papers
%%% January 27, 2015 - first edition
%%% February 2, 2015 - add Yelton, Badhrees
%%% February 5, 2015 - add Babu, Prasanth
%%% Non-responding authors or those who said NO are commented out.
%%% ====================================================================
%%% Click the RELOAD button on your web browser to see the updated file.
%%% ====================================================================
%%% Use \input{author} to insert this material into your latex file.
%%%%% Force institutions to appear in alphabetical order when typeset.
%%% Paper:
%%% Journal:  2015 Conference Papers
%%% January 27, 2015 - first edition
%%% February 2, 2015 - add Yelton, Badhrees
%%% February 5, 2015 - add Babu, Prasanth
%%% Non-responding authors or those who said NO are commented out.
%%% ====================================================================
%%% Click the RELOAD button on your web browser to see the updated file.
%%% ====================================================================
%%% Use \input{author} to insert this material into your latex file.
%%%%% Force institutions to appear in alphabetical order when typeset.
\noaffiliation
\affiliation{University of the Basque Country UPV/EHU, 48080 Bilbao}
\affiliation{Beihang University, Beijing 100191}
\affiliation{University of Bonn, 53115 Bonn}
\affiliation{Budker Institute of Nuclear Physics SB RAS and Novosibirsk State University, Novosibirsk 630090}
\affiliation{Faculty of Mathematics and Physics, Charles University, 121 16 Prague}
\affiliation{Chiba University, Chiba 263-8522}
\affiliation{Chonnam National University, Kwangju 660-701}
\affiliation{University of Cincinnati, Cincinnati, Ohio 45221}
\affiliation{Deutsches Elektronen--Synchrotron, 22607 Hamburg}
\affiliation{University of Florida, Gainesville, Florida 32611}
\affiliation{Department of Physics, Fu Jen Catholic University, Taipei 24205}
\affiliation{Justus-Liebig-Universit\"at Gie\ss{}en, 35392 Gie\ss{}en}
\affiliation{Gifu University, Gifu 501-1193}
\affiliation{II. Physikalisches Institut, Georg-August-Universit\"at G\"ottingen, 37073 G\"ottingen}
\affiliation{The Graduate University for Advanced Studies, Hayama 240-0193}
\affiliation{Gyeongsang National University, Chinju 660-701}
\affiliation{Hanyang University, Seoul 133-791}
\affiliation{University of Hawaii, Honolulu, Hawaii 96822}
\affiliation{High Energy Accelerator Research Organization (KEK), Tsukuba 305-0801}
\affiliation{Hiroshima Institute of Technology, Hiroshima 731-5193}
\affiliation{IKERBASQUE, Basque Foundation for Science, 48013 Bilbao}
\affiliation{University of Illinois at Urbana-Champaign, Urbana, Illinois 61801}
\affiliation{Indian Institute of Technology Bhubaneswar, Satya Nagar 751007}
\affiliation{Indian Institute of Technology Guwahati, Assam 781039}
\affiliation{Indian Institute of Technology Madras, Chennai 600036}
\affiliation{Indiana University, Bloomington, Indiana 47408}
\affiliation{Institute of High Energy Physics, Chinese Academy of Sciences, Beijing 100049}
\affiliation{Institute of High Energy Physics, Vienna 1050}
\affiliation{Institute for High Energy Physics, Protvino 142281}
\affiliation{Institute of Mathematical Sciences, Chennai 600113}
\affiliation{INFN - Sezione di Torino, 10125 Torino}
\affiliation{Institute for Theoretical and Experimental Physics, Moscow 117218}
\affiliation{J. Stefan Institute, 1000 Ljubljana}
\affiliation{Kanagawa University, Yokohama 221-8686}
\affiliation{Institut f\"ur Experimentelle Kernphysik, Karlsruher Institut f\"ur Technologie, 76131 Karlsruhe}
\affiliation{Kavli Institute for the Physics and Mathematics of the Universe (WPI), University of Tokyo, Kashiwa 277-8583}
\affiliation{King Abdulaziz City for Science and Technology, Riyadh 11442}
\affiliation{Department of Physics, Faculty of Science, King Abdulaziz University, Jeddah 21589}
\affiliation{Korea Institute of Science and Technology Information, Daejeon 305-806}
\affiliation{Korea University, Seoul 136-713}
\affiliation{Kyoto University, Kyoto 606-8502}
\affiliation{Kyungpook National University, Daegu 702-701}
\affiliation{\'Ecole Polytechnique F\'ed\'erale de Lausanne (EPFL), Lausanne 1015}
\affiliation{Faculty of Mathematics and Physics, University of Ljubljana, 1000 Ljubljana}
\affiliation{Luther College, Decorah, Iowa 52101}
\affiliation{University of Maribor, 2000 Maribor}
\affiliation{Max-Planck-Institut f\"ur Physik, 80805 M\"unchen}
\affiliation{School of Physics, University of Melbourne, Victoria 3010}
\affiliation{Moscow Physical Engineering Institute, Moscow 115409}
\affiliation{Moscow Institute of Physics and Technology, Moscow Region 141700}
\affiliation{Graduate School of Science, Nagoya University, Nagoya 464-8602}
\affiliation{Kobayashi-Maskawa Institute, Nagoya University, Nagoya 464-8602}
\affiliation{Nara University of Education, Nara 630-8528}
\affiliation{Nara Women's University, Nara 630-8506}
\affiliation{National Central University, Chung-li 32054}
\affiliation{National United University, Miao Li 36003}
\affiliation{Department of Physics, National Taiwan University, Taipei 10617}
\affiliation{H. Niewodniczanski Institute of Nuclear Physics, Krakow 31-342}
\affiliation{Nippon Dental University, Niigata 951-8580}
\affiliation{Niigata University, Niigata 950-2181}
\affiliation{University of Nova Gorica, 5000 Nova Gorica}
\affiliation{Osaka City University, Osaka 558-8585}
\affiliation{Osaka University, Osaka 565-0871}
\affiliation{Pacific Northwest National Laboratory, Richland, Washington 99352}
\affiliation{Panjab University, Chandigarh 160014}
\affiliation{Peking University, Beijing 100871}
\affiliation{University of Pittsburgh, Pittsburgh, Pennsylvania 15260}
\affiliation{Punjab Agricultural University, Ludhiana 141004}
\affiliation{Research Center for Electron Photon Science, Tohoku University, Sendai 980-8578}
\affiliation{Research Center for Nuclear Physics, Osaka University, Osaka 567-0047}
\affiliation{RIKEN BNL Research Center, Upton, New York 11973}
\affiliation{Saga University, Saga 840-8502}
\affiliation{University of Science and Technology of China, Hefei 230026}
\affiliation{Seoul National University, Seoul 151-742}
\affiliation{Shinshu University, Nagano 390-8621}
\affiliation{Soongsil University, Seoul 156-743}
\affiliation{Sungkyunkwan University, Suwon 440-746}
\affiliation{School of Physics, University of Sydney, NSW 2006}
\affiliation{Department of Physics, Faculty of Science, University of Tabuk, Tabuk 71451}
\affiliation{Tata Institute of Fundamental Research, Mumbai 400005}
\affiliation{Excellence Cluster Universe, Technische Universit\"at M\"unchen, 85748 Garching}
\affiliation{Toho University, Funabashi 274-8510}
\affiliation{Tohoku Gakuin University, Tagajo 985-8537}
\affiliation{Tohoku University, Sendai 980-8578}
\affiliation{Department of Physics, University of Tokyo, Tokyo 113-0033}
\affiliation{Tokyo Institute of Technology, Tokyo 152-8550}
\affiliation{Tokyo Metropolitan University, Tokyo 192-0397}
\affiliation{Tokyo University of Agriculture and Technology, Tokyo 184-8588}
\affiliation{University of Torino, 10124 Torino}
\affiliation{Toyama National College of Maritime Technology, Toyama 933-0293}
\affiliation{Utkal University, Bhubaneswar 751004}
\affiliation{CNP, Virginia Polytechnic Institute and State University, Blacksburg, Virginia 24061}
\affiliation{Wayne State University, Detroit, Michigan 48202}
\affiliation{Yamagata University, Yamagata 990-8560}
\affiliation{Yonsei University, Seoul 120-749}
  \author{A.~Abdesselam}\affiliation{Department of Physics, Faculty of Science, University of Tabuk, Tabuk 71451} % Tabuk
  \author{I.~Adachi}\affiliation{High Energy Accelerator Research Organization (KEK), Tsukuba 305-0801}\affiliation{The Graduate University for Advanced Studies, Hayama 240-0193} % KEK
  \author{K.~Adamczyk}\affiliation{H. Niewodniczanski Institute of Nuclear Physics, Krakow 31-342} % Krakow
  \author{H.~Aihara}\affiliation{Department of Physics, University of Tokyo, Tokyo 113-0033} % Tokyo
  \author{S.~Al~Said}\affiliation{Department of Physics, Faculty of Science, University of Tabuk, Tabuk 71451}\affiliation{Department of Physics, Faculty of Science, King Abdulaziz University, Jeddah 21589} % Tabuk
  \author{K.~Arinstein}\affiliation{Budker Institute of Nuclear Physics SB RAS and Novosibirsk State University, Novosibirsk 630090} % BINP
  \author{Y.~Arita}\affiliation{Graduate School of Science, Nagoya University, Nagoya 464-8602} % Nagoya
  \author{D.~M.~Asner}\affiliation{Pacific Northwest National Laboratory, Richland, Washington 99352} % PNNL
  \author{T.~Aso}\affiliation{Toyama National College of Maritime Technology, Toyama 933-0293} % Toyama
  \author{V.~Aulchenko}\affiliation{Budker Institute of Nuclear Physics SB RAS and Novosibirsk State University, Novosibirsk 630090} % BINP
% ...
% BINP
\collaboration{The Belle Collaboration}

\begin{abstract}

We report a measurement of the amplitude ratio $r_S$ of $B^0 \to D^0K^{*0}$ and $B^0 \to \bar{D^0}K^{*0}$ decays 
with a model-independent Dalitz plot analysis using $D\to K_S^0\pi^+\pi^-$ decays.
Using the full data sample of $772\times10^6$ $B\bar{B}$ pairs collected 
at the $\Upsilon(4S)$ resonance with the Belle detector at KEKB accelerator the upper limit is
$r_S  <  0.87$ at the 68~\%  confidence level.
This result is the first measurement of $r_S$ with a model-independent Dalitz analysis, and is consistent
with results from other analyses.
The value of $r_S$ indicates the sensitivity of the decay to $\phi_3$ because the statistical uncertainty is proportional to $1/r_S$. 
The $r_S$ result is obtained from observables ($x_\pm$, $y_\pm$)
\begin{eqnarray}
x_- &=& +0.4 ^{+1.0 +0.0}_{-0.6 -0.1} \pm0.0	\nonumber \\
y_- &=& -0.6 ^{+0.8 +0.1}_{-1.0 -0.0} \pm0.1	\nonumber \\
x_+ &=& +0.1 ^{+0.7 +0.0}_{-0.4 -0.1} \pm0.1	\nonumber \\
y_+ &=& +0.3 ^{+0.5 +0.0}_{-0.8 -0.1} \pm0.1 , \nonumber
\end{eqnarray}
where $x_\pm = r_S \cos(\delta_S \pm \phi_3)$, $y_\pm = r_S \sin(\delta_S \pm \phi_3)$ and
$\phi_3~(\delta_S)$ are the weak (strong) phase difference between $B^0 \to D^0K^{*0}$ and $B^0 \to \bar{D^0}K^{*0}$.
The first uncertainty is statistical, the second is the experimental systematic and the third is the systematic due to the uncertainties on $c_i$ and $s_i$ parameters measured by CLEO.

\end{abstract}


\maketitle

\tighten

{\renewcommand{\thefootnote}{\fnsymbol{footnote}}}
\setcounter{footnote}{0}

\section{INTRODUCTION}

Determination of the parameters of the standard model (SM)
plays an important role in the search for new physics.
In the SM,
the Cabibbo-Kobayashi-Maskawa (CKM) matrix~\cite{CKM} gives a successful description of current measurements of $CP$ violation.
$CP$-violating parameters $\phi_1$, $\phi_2$ and $\phi_3$ are the three angles of a CKM unitarity triangle, of which
$\phi_3 \equiv \arg{(-V_{ud}{V_{ub}}^*/V_{cd}{V_{cb}}^*)}$
is the least accurately determined. %~\cite{phi12}.
In the usual quark-phase convention, where large complex phases appear only in $V_{ub}$ and $V_{td}$~\cite{Wolfenstein},
the measurement of $\phi_3$ is equivalent to the extraction of the phase of $V_{ub}$ relative to the phases of the other CKM matrix elements except $V_{td}$.
To date, $\phi_3$ measurements have been made mainly with $B$ meson decays into $D^{(*)}K^{(*)}$ final states
~\cite{GLW_Belle_result, Dalitz_Belle_result, GLW_CDF_result, GLW_BaBar_result, ADS_BaBar_result, Dalitz_BaBar_result, ADS_CDF_result, ADS_Belle_result, LHCb_result, BaBar_DKst0},
which exploit the interference between the $\bar{D}^{(*)0}$ and $D^{(*)0}$ decaying into a common final state.
In particular, Dalitz plot analyses of $D$ decays in $B^\pm\to D^{(*)}K^{(*)\pm}$ processes are the most sensitive to $\phi_3$ at $e^+e^-$ B-factories.
The Dalitz plot analysis technique for the measurement of $\phi_3$ was proposed in Ref.~\cite{GGSZ, GGSZ_2}. 
Belle reported the first $\phi_3$ measurement with model-independent Dalitz analysis technique~\cite{Antn_Mod_Ind_Dalitz}.

\begin{figure}[ht]
  \begin{center}
  \leavevmode
   \subfigure
%    {\includegraphics[width=0.41\textwidth, bb=0 0 374 122, clip]{1.pdf}}
    {\includegraphics[width=0.41\textwidth, bb=0 0 374 122, clip]{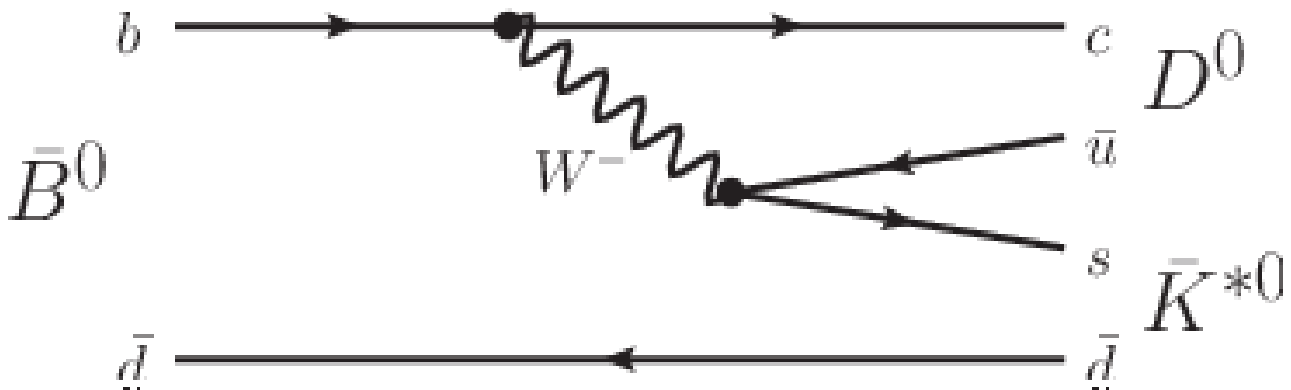}}
    \hspace{6mm}
    \subfigure
%    {\includegraphics[width=0.41\textwidth, bb=0 0 374 122]{2.pdf}}
    {\includegraphics[width=0.41\textwidth, bb=0 0 374 122]{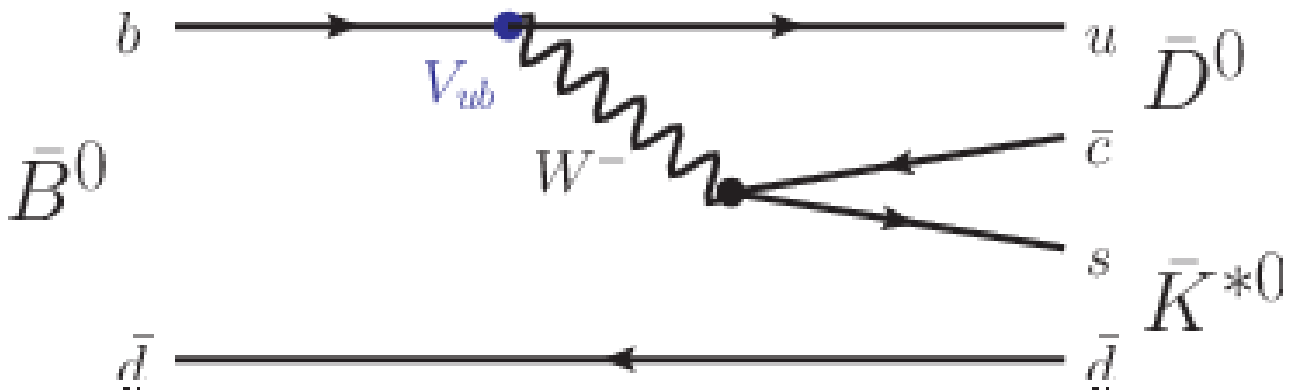}}
    \caption{Diagrams for the $\bar{B}^0 \to \bar{D}^0 \bar{K}^{*0}$ and $\bar{B}^0 \to D^0 \bar{K}^{*0}$ decays.}
    \label{fig:Dgrm_DKst0}
  \end{center}
\end{figure}

In this paper,
we present the first measurement of the amplitude ratio of $B^0 \to D^0K^{*0}$ and $B^0 \to \bar{D^0}K^{*0}$ decays
with a model-independent Dalitz plot analysis, which is essential for the determination of $\phi_3$ from this channel.
We reconstruct $B^0\to DK^{*0}$, with $K^{*0}\to K^+\pi^-$ (charge conjugate processes are implied throughout the paper and $K^{*0}$ refers to $K^{*}(892)^{0}$).
Here the flavor of the $B$ meson is identified by the kaon charge.
Neutral $D$ mesons are reconstructed in the $K_S^0\pi^+\pi^-$ decay mode.
The final states we reconstructed can be reached through $b\to c$ and $b\to u$ processes with the diagrams shown in Fig.~\ref{fig:Dgrm_DKst0}.

Since the $K^*$ has a large natural width of $50~{\rm MeV}$, 
we consider interference between the signal $B \to D K^*$ and $B \to D K \pi$ where $K \pi$
arises from a non-resonant decay or through
other kaonic resonances. In this analysis we use the variables $r_S$, $k$, and $\delta_S$ to parametrize the strong dynamics of the decay. 
These parameters are defined as \cite{ADS_2}
\begin{eqnarray}
  r_S^2 &\equiv& \frac{\Gamma(B^0\to D^0K^+\pi^-)}{\Gamma(B^0\to \bar{D}^0K^+\pi^-)} = \frac{\int dp A_{b\to u}^2(p)}{\int dp A_{b\to c}^2(p)}, \\
  k\mathrm{e}^{i\delta_S} &\equiv& \frac{\int dp A_{b\to c}(p)A_{b\to u}(p)\mathrm{e}^{i\delta(p)}}{\sqrt{\int dp A_{b\to c}^2(p) \int dp A_{b\to u}^2(p)}},
  \label{eq:rS_delS_k}
\end{eqnarray}
where the integration is over the $B^0\to DK^+\pi^-$ Dalitz plot region corresponding to the $K^{*0}$ resonance.
Here $A_{b\to c}(A_{b\to u})(p)$ are the magnitudes of the amplitudes for the $b\to c(u)$ transitions,
and $\delta(p)$ is the relative strong phase, where the variable $p$ indicates the position within the $DK^+\pi^-$ Dalitz plot.
If the $B^0$ decay can be considered as a $DK^{*0}$ two-body decay,
$r_S$ becomes the ratio of the amplitudes for $b\to u$ and $b\to c$ and $k$ becomes $1$.
According to a simulation study using a Dalitz model based on the measurements~\cite{BaBar_k},
the value of $k$ is $0.95\pm0.03$ within the phase space of $DK^{*0}$ resonance.
The value of $r_S$ is expected to be around $0.4$,
which naively corresponds to $\mid V_{ub}V^*_{cs}\mid / \mid V_{cb}V^*_{us}\mid$,
but also depends on strong interaction effects.

\section{THE MODEL-INDEPENDENT DALITZ PLOT ANALYSIS TECHNIQUE}

The amplitude of the $B^0\to DK^{*0}$, $D\to K_S^0\pi^+\pi^-$
decay is a superposition of the $B^0\to \bar{D}^0K^{*0}$ and $B^0\to D^0K^{*0}$ amplitudes
\begin{eqnarray}
A_B(m_+^2,m_-^2) = \bar{A} + r_S e^{i(\delta_S + \phi_3)}A,
	\label{eq:Amp_Inf}
\end{eqnarray}
where $m_+^2$ and $m_-^2$ are the squared invariant masses of $K_S^0\pi^+$ and $K_S^0\pi^-$ combinations, respectively,
$\bar{A} = \bar{A}(m_+^2,m_-^2)$
is the amplitude of the $\bar{D}^0\to K_S^0\pi^+\pi^-$ decay and
$A = A(m_+^2,m_-^2)$
is the amplitude of the $D^0\to K_S^0\pi^+\pi^-$ decay.
In the case of $CP$ conservation in the $D$ decay
$A(m_+^2,m_-^2) = \bar{A}(m_-^2,m_+^2)$.
The Dalitz plot density of the $D$ decay from $B^0\to DK^{*0}$ is given by
\begin{eqnarray}
P_B = \mid A_B \mid^2 = \mid \bar{A} + r_S e^{i(\delta_S + \phi_3)}A \mid^2 = \bar{P} + r_S^2P + 2k\sqrt{P\bar{P}}(x_+C + y_+S),
	\label{eq:Dcy_inf}
\end{eqnarray}
where $P(m_+^2,m_-^2) = \mid A \mid^2$, $\bar{P}(m_+^2,m_-^2) = \mid \bar{A} \mid^2$, and
\begin{eqnarray}
x_+ =r_S\cos(\delta_S + \phi_3), \hspace{20pt} y_+ = r_S\sin(\delta_S + \phi_3).
	\label{eq:x}
\end{eqnarray}
Functions
$C = C(m_+^2,m_-^2)$ and
$S = S(m_+^2,m_-^2)$
are the cosine and sine of the strong-phase difference
$\delta_D(m_+^2,m_-^2) = \arg \bar{A} - \arg A$ between the $\bar{D}^0\to K_S^0\pi^+\pi^-$ and $D^0\to K_S^0\pi^+\pi^-$ amplitudes. Here, we have used the definition of $k$ given in Eq.~(\ref{eq:rS_delS_k}).
The equations for the charge-conjugate mode $\bar{B}^0\to D\bar{K}^{*0}$ are obtained with the substitution $\phi_3 \leftrightarrow -\phi_3$ and $A \leftrightarrow \bar{A}$;
the corresponding parameters that depend on the $\bar{B}^0$ decay amplitude are:
\begin{eqnarray}
x_- =r_S\cos(\delta_S - \phi_3), \hspace{20pt} y_- = r_S\sin(\delta_S - \phi_3).
	\label{eq:x_y}
\end{eqnarray}
Using $B^0$ and $\bar{B}^0$ decays, one can obtain $r_S$, $\phi_3$ and $\delta_S$ separately.

Up to this point, the description of the model-dependent and model-independent techniques is the same.
The model-dependent analysis deals directly with the Dalitz plot density,
and the functions $C$ and $S$ are obtained from a model based upon a fit to the $D^0\to K_S^0\pi^+\pi^-$ amplitude.
In the model-independent approach~\cite{MIDalitz_phi3, MIDalitz_phi3_2}, the Dalitz plot is divided into $2{\cal N}$ bins symmetric under the exchange $m_-^2 \leftrightarrow m_+^2$.
The bin index $i$ ranges from $-{\cal N}$ to ${\cal N}$ (excluding $0$);
the exchange $m_-^2 \leftrightarrow m_+^2$ corresponds to the exchange $i \leftrightarrow -i$.
The expected number of events in bin $i$ of the Dalitz plot of the $D$ meson from $B^0\to DK^{*0}$ is
\begin{eqnarray}
N^{\pm}_i = h_B \left[ K_{\pm i} + r_S^2K_{\mp i} + 2k\sqrt{K_i K_{-i}}(x_{\pm}c_i \pm y_{\pm}s_i) \right],
	\label{eq:N_i}
\end{eqnarray}
where $N^{+(-)}$ stands for the number of $B^0(\bar{B}^0)$ meson decays, $h_B$ is a normalization constant and $K_i$ is the number of events in the $i^{\mathrm{th}}$ bin of the $K_S^0\pi^+\pi^-$ Dalitz plot of the $D$ meson.
The values of $K_i$ are measured from a sample of flavor-tagged $D^0$ mesons obtained by reconstructing $D^{*\pm}\to D\pi^\pm$ decays.
The terms $c_i$ and $s_i$ are the amplitude-weighted average of the functions $C$ and $S$ over the bin region:
\begin{eqnarray}
c_i = \frac{\int_{{\cal D}_i} \mid A\mid\mid \bar{A}\mid Cd{\cal D}}
{\sqrt{\int_{{\cal D}_i} \mid A\mid^2 d{\cal D}\int_{{\cal D}_i} \mid \bar{A}\mid^2 d{\cal D}}}.
	\label{eq:c_i}
\end{eqnarray}
Here $\cal D$ represents the Dalitz plot plane and ${\cal D}_i$ is the bin region over which the integration is performed.
The terms $s_i$ are defined similarly with $C$ substituted by $S$.
The absence of $CP$ violation in the $D$ decay implies $c_i = c_{-i}$ and $s_i = -s_{-i}$.
The values of $c_i$ and $s_i$ can be measured using quantum correlated $D$ pairs
produced at charm-factory experiments operating at the threshold for $D\bar{D}$ pair production.
The CLEO Collaboration has reported $c_i$ and $s_i$ values \cite{cisi_CLEO_1, cisi_CLEO_2}.
Once the $c_i$ and $s_i$ measurements are included, the set of relations defined by Eq.~(\ref{eq:N_i}) contain
only three free parameters ($x$, $y$, and $h_B$) for each $B^0$,
and can be solved using a maximum likelihood method to extract the values of $\phi_3$, $\delta_{S}$ and $r_S$.
We have neglected charm-mixing effects in $D$ decays from both the $B^0\to DK^{*0}$ process and in the quantum correlated $D\bar{D}$ production.

In principle, the set of relations defined by Eq.~(\ref{eq:N_i}) can be solved without external constraints on $c_i$ and $s_i$ for ${\cal N} \geq 2$.
However, due to the small value of $r_S$, there is very little sensitivity to the $c_i$ and $s_i$ parameters in $B^0\to DK^{*0}$ decays,
which results in a reduction in the precision on the other parameters.

\section{ANALYSIS PROCEDURE}

Equation~(\ref{eq:N_i}) is the key relation used in the analysis,
but it only holds if there is no background,
no non-uniformity in the Dalitz plot acceptance and no crossfeed between bins.
(Cross-feed is due to invariant-mass resolution and radiative corrections.)
In this section we outline the procedures to account for the acceptance and crossfeed.

First we discuss the effect of the variation of efficiency profile over the Dalitz plane.
We note that Eq.~(\ref{eq:Dcy_inf}) does not change under the transformation $P\to \epsilon P$ when the efficiency profile
$\epsilon(m_+^2, m_-^2)$ is symmetric:
$\epsilon(m_+^2, m_-^2) = \epsilon(m_-^2, m_+^2)$.
The effect of non-uniform efficiency over the Dalitz plot cancels out
when using a flavor-tagged $D$ sample with kinematic properties that are similar to the sample from the signal $B$ decay.
This approach allows for the removal of the systematic uncertainty associated
with the possible inaccuracy of the detector acceptance description in the Monte Carlo (MC) simulation.
With the efficiency taken into account (that is in general non-uniform across the bin region), the number of events detected is:
\begin{eqnarray}
N' = \int p({\cal D})\epsilon({\cal D})d{\cal D} .
	\label{eq:N_i_eff}
\end{eqnarray}
Clearly, the efficiency does not factorize. One can use an efficiency averaged over the bin then correct for it in the analysis:
\begin{eqnarray}
\bar{\epsilon}_i = \frac{N_i'}{N_i} = \frac{\int p({\cal D})\epsilon({\cal D})d{\cal D}}{\int p({\cal D})d{\cal D}}
	\label{eq:eff}
\end{eqnarray}
The averaged efficiency $\bar{\epsilon}_i$ can be determined from MC.
The assumption that the efficiency profile depends only on the $D$ momentum is tested using MC simulation,
and the remaining difference is treated as a systematic uncertainty.
This correction cannot be performed in a completely model-independent way,
since the correction terms include the amplitude  variation inside the bin.
Fortunately, calculations using the Belle $D\to K_S^0\pi^+\pi^-$ model ~\cite{Dalitz_mod_Belle} show that this correction is negligible even for very large non-uniformity of the efficiency profile.

There are two sources of cross-feed: momentum resolution and flavor misidentification.
Momentum resolution leads to migration of events between the bins.
In the binned approach, this effect can be corrected in a non-parametric way.
The migration can be described by a linear transformation of the number of events in each bin
\begin{eqnarray}
N_i' = \sum \alpha_{ik}N_{k},
\end{eqnarray}
where $N_k$ is the number of events that bin $k$ would contain without the cross-feed, and $N_i'$ is the reconstructed number of events in bin $i$.
The crossfeed matrix $\alpha_{ik}$ is nearly a unit matrix: $\alpha_{ik} \ll 1$ for $i \neq k$.
The matrix is obtained from a signal MC simulation generated with the amplitude model reported in Ref.~\cite{Dalitz_mod_Belle}.
Most of the off-diagonal elements are null, only a few have values $\alpha_{ik} \le 0.04$. 
In the case of a $D\to K_S^0\pi^+\pi^-$ decay from a $B$, the cross-feed depends on the parameters $x$ and $y$.
However, this is a minor correction to an already small effect due to cross-feed; therefore it is neglected.

The final effect to be considered are events in which the $B$ flavor is misidentified.
Double mis-identification in $K^{*0}$ reconstruction from $K^+\pi^-$, leads to migration of events between $N_i^+ \leftrightarrow N_{-i}^-$,
due to assignment of the wrong flavor to the $B$  candidate. If the fraction of double mis-identified events is $\beta$, the number of events in each bin can be  written as
\begin{eqnarray}
N_i^{\prime \pm} = N_i^\pm + \beta N_{-i}^\mp.
\end{eqnarray}
The value of $\beta$ is obtained from  MC data and is found to be $(0.119 \pm 0.007)\%$.
Therefore, in this analysis the effect of flavor misidentification is neglected.

\subsection{Fit Procedure}

We can fit the data distribution in each bin separately,
with the number of signal and background events as free parameters.
The values of $N_i$ found can be used in Eq.~(\ref{eq:N_i}) to obtain the
parameters ($x_\pm$, $y_\pm$).
This is accomplished by minimizing a negative logarithmic likelihood of the form
\begin{eqnarray}
-2\log{\cal L}(x, y) = -2\sum_i \log p(\left< N_i \right>(x, y), N_i, \sigma_{N_i}),
\end{eqnarray}
where $\left< N_i \right>$ is the expected number of events in the bin $i$ obtained from Eq.~(\ref{eq:N_i}).
Here, $N_i$ and $\sigma_{N_i}$ are the observed number of events in data and the uncertainty on $N_i$, respectively.
If the probability density function (PDF) $p$ is Gaussian, this procedure is equivalent to a $\chi^2$ fit;
however, the assumption of the Gaussian distribution may introduce a bias in the case of low statistics in certain bins.

The procedure described above does not make any assumptions on the Dalitz distribution of the background events,
since the fits in each bin are independent.
Thus there is no associated systematic uncertainty.
However, in the case of a small number of events and many background components this can be a limiting factor.
Our approach is to use the combined fit with a common likelihood for all bins.
The background contribution has to be accounted for in the calculation of the values $N_i$.
Statistically the most effective way of calculating the number of signal events is to perform,
in each bin $i$ of the Dalitz plot, an unbinned fit in the variables used to distinguish the signal from the background.
In this analysis, we obtain the $CP$ violating parameters ($x_\pm$, $y_\pm$) from the data from a combined fit in bins.
The relative numbers of background events in the bins of such a fit can be constrained externally from MC samples.
In the combined fit, the expected numbers of events $\left< N_i \right>$ as functions of ($x$, $y$) can be included in the likelihood.
Thus the variables ($x$, $y$) become free parameters of the combined likelihood fit,
and the assumption that the number of signal events has a Gaussian distribution is not needed.

\section{EVENT RECONSTRUCTION AND SELECTION}

This analysis is based on a data sample that contains $772~\times 10^6~B\bar{B}$ pairs,
collected  with the Belle detector at the KEKB asymmetric-energy $e^+e^-$ ($3.5$ on $8~{\rm GeV}$) collider~\cite{KEKB} operating at the $\Upsilon(4S)$ resonance.
The Belle detector is a large-solid-angle magnetic spectrometer that consists of a silicon vertex detector,
a $50$-layer central drift chamber (CDC),
an array of aerogel threshold Cherenkov counters (ACC),
a barrel-like arrangement of time-of-flight scintillation counters (TOF),
and an electromagnetic calorimeter comprised of CsI(Tl) crystals located inside a superconducting solenoid coil that provides a $1.5~{\rm T}$ magnetic field.
An iron flux-return located outside of the coil is instrumented to detect $K_L^0$ mesons and to identify muons.
The detector is described in detail elsewhere~\cite{Belle}.

We reconstructed $B^0\to DK^{*0}$ events with $K^{*0}\to K^+\pi^-$ and $D\to K_S^0\pi^+\pi^-$.
The event selection, described below, is developed from studies of off-resonance data and MC simulated events.

The $K_S^0$ candidates are identified using the output of a neural network. Inputs to the network for a pair of oppositely-charged pions are the invariant mass, 20 kinematic parameters and particle identification (PID) information  from the ACC, TOF and the ionization energy loss in the CDC.
The $K_S^0$ selection has a simulated purity of $92.2\%$ and an efficiency of $75.1\%$.
Charged kaon and pion candidates are identified using PID information. The efficiency is $85{\rm -}95\%$ and the probability of misidentification is $10{\rm -}20\%$ depending upon the momentum of the hadrons as obtained using dedicated data control samples.
We reconstruct neutral $D$ mesons by combining a $K_S^0$ candidate with a pair of oppositely-charged pion candidates.
We require that the invariant mass is within $\pm15~{\rm MeV}/c^2$ ($\pm 3 \sigma$) of the nominal $D^0$ mass.
$K^{*0}$ candidates are reconstructed from $K^+\pi^-$ pairs.
We require that the invariant mass is within $\pm$50~${\rm MeV}/c^2$ of the nominal $K^{*0}$ mass.
We combine $D$ and $K^{*0}$ candidates to form $B^0$ mesons.
Candidate events are identified by the energy difference $\Delta E \equiv \sum_{i}E_i - E_{\mathrm b}$
and the beam-constrained mass $M_{\rm bc} \equiv \sqrt{E_{\mathrm b}^2 - \mid\sum_{i}\vec{p}_i\mid^2}$,
where $E_{\mathrm b}$ is the beam energy and $\vec{p}_i$ and $E_i$ are the momenta and energies, respectively,
of the $B^0$ meson decay products in the $e^+e^-$ center-of-mass (CM) frame.
We select events with $5.2~{\rm GeV}/c^2 < M_{\rm bc} < 5.29~{\rm GeV}/c^2$ and $-0.1~{\rm GeV} < \Delta E < 0.15~{\rm GeV}$.

Among other $B$ decays, the most serious background is from $\bar{B}^0\to [\bar{K}^{*0}\pi^+]_{D^+} [K^0\pi^-]_{K^{*-}}$.
This decay produces the same final state as the $B^0\to DK^{*0}$ signal.
To suppress this background, we exclude candidates for which the invariant mass of the $K^{*0} \pi^+$ system is within $\pm4~{\rm MeV}/c^2$ of the nominal $D^+$ mass. This criteria leads to negligible contamination from $\bar{B}^0\to [\bar{K}^{*0}\pi^+]_{D^+} [K^0\pi^-]_{K^{*-}}$ and a relative loss in the signal efficiency of $0.6\%$.

Large combinatorial background of true $D^0$ and random $K^+$ and $\pi^-$ combinations from the $e^+e^-\to c\bar{c}$ process and other $B\bar{B}$ decays
is reduced if $D^0$ candidates that are a decay product of $D^{*+} \to D^0 \pi^+$ are eliminated.
We use the mass difference $\Delta M$ between the $[K_S^0\pi^+\pi^-]_D \pi^+$ and $[K_S^0\pi^+\pi^-]_D$ systems.
If $\Delta M > 0.15~{\rm GeV}/c^2$ for any additional $\pi^+$ candidate not used in the $B$ candidate reconstruction, the event is retained.
This requirement removes $19\%$ of $c\bar{c}$ background and $11\%$ of $B\bar{B}$ background according to MC simulation.
The relative loss in signal efficiency is $5.5\%$

In the rare case where there are multiple candidates in an event,
the candidate with $M_{\rm bc}$ closest to its nominal value is chosen.
The relative loss in signal efficiency is $0.8 \%$.

To discriminate the large combinatorial background dominated by the two-jet-like $e^+e^-\rightarrow q\bar{q}$ continuum process,
where $q$ indicates $u$, $d$, $s$ or $c$, a multivariate analysis is performed using the 12 variables introduced in the Table~\ref{tab:paras}.
\begin{table}
\begin{center}
\begin{tabular}{| c | l |}
\hline
1 & \parbox{380pt}{\strut{} 
Fisher discriminants based on modified Fox-Wolfram moments~\cite{SFW}.
\strut} \\
\hline
2 & \parbox{380pt}{\strut{}
 The angle in the CM frame between the thrust axes of the $B$ decay
\\ and that of remaining particles.
\strut} \\
\hline
3 & \parbox{380pt}{\strut{}
The signed difference of the vertices between the $B$ candidate\\ and the remaining charged tracks.
\strut} \\
\hline
4 & \parbox{380pt}{\strut{}
The distance of closest approach between the trajectories of the $K^{*}$ and $D$ candidates.
\strut} \\
\hline
5 & \parbox{380pt}{\strut{}
The expected flavor dilution factor described in Ref.~\cite{TaggingNIM}.
\strut} \\
\hline
6 & \parbox{380pt}{\strut{}
The angle $\theta$ between the $B$ meson momentum direction and the beam axis in the CM frame.
\strut} \\
\hline
7 & \parbox{380pt}{\strut{}
The angle between the $D$ and $\Upsilon(4S)$ directions in the rest frame of the $B$ candidate.
\strut} \\
\hline
8 & \parbox{380pt}{\strut{}
The projection on to the $e^+e^-$ beam direction of the sphericity vector\\ with the largest eigenvalue.
\strut} \\
\hline
9 & \parbox{380pt}{\strut{}
The angle of the sphericity vector of the signal with respect to that of the remaining particles\\ with the largest eigenvalue.
\strut} \\
\hline
10 & \parbox{380pt}{\strut{}
The angle of the sphericity vector of the signal with respect to the remaining particles\\ with the second largest eigenvalue.
\strut} \\
\hline
11 & \parbox{380pt}{\strut{}
The angle of sphericity vector from the signal with respect to the remaining particles\\ with the smallest eigenvalue.
\strut} \\
\hline
12 & \parbox{380pt}{\strut{}
The magnitude of the thrust of the particles not used to reconstruct the signal.
\strut} \\
\hline
\end{tabular}
\caption{12 variables for $q\bar{q}$ suppression.}
\label{tab:paras}
\end{center}
\end{table}
To effectively combine these 12 variables,
we employ the NeuroBayes neural network package \cite{NB}.
The NeuroBayes output is denoted as $C_{\rm NB}$ and lies within the range  [$-1$, 1].
Events with $C_{\rm NB} \sim 1$ are signal-like and events with $C_{\rm NB} \sim -1$ are $q\bar{q}$-like.
The training of the neural network is performed using signal and $q\bar{q}$ MC samples.
The $C_{\rm NB}$ distribution of signal events peaks at $|C_{\rm NB}|\sim 1$ and is therefore difficult to represent with a simple analytic function.
However, the transformed variable
\begin{eqnarray}
  C'_{\rm NB} &=& \ln \frac{C_{\rm NB}-{C_{{\rm NB,} {\rm low}}}}{C_{{\rm NB,} {\rm high}}-C_{\rm NB}}\ ,
\end{eqnarray}
where $C_{{\rm NB,} {\rm low}} = -0.6$ and $C_{{\rm NB,} {\rm high}} = 0.9992$,
has a distribution that can be modelled by a Gaussian for signal as well as background.
The events with $C_{\rm NB} < -0.6$ are rejected.

The number of signal events is obtained by fitting the three-dimensional distribution of variables $M_{\rm bc}$, $\Delta E$, and $C'_{\rm NB}$
using the extended maximum likelihood method.
We prepare three-dimensional PDFs for each component by forming the product of one-dimensional PDFs for $\Delta E$, $M_{\rm bc}$ and $C'_{\rm NB}$,
since the correlations among the variables
are found to be small.

Backgrounds are divided into the following components:
\begin{itemize}
\item Combinatorial background from $q\bar{q}$ background events.
\item $B\bar{B}$ background, in which the tracks forming the $B^0\to DK^{*0}$ candidate come from decays of both $B$ mesons in the event.
The number of possible $B$ decay combinations that contribute to this background is large,
therefore both the Dalitz distribution and fit parameters distribution are quite smooth.
In this analysis, $B\bar{B}$ backgrounds are further subdivided into two components:
events reconstructed with a true $D\to K^{0}_{S}\pi^{+}\pi^{-}$ decay, referred to as $D$ true, and those reconstructed with a combinatorial $D$ candidate, referred to as $D$ fake.
\item Peaking $B\bar{B}$ background, in which all tracks forming the $B^0\to DK^{*0}$ candidate come from the same $B$ meson.
This background has two types: events with one pion misidentified as a kaon, such as $D^0 [\pi^+\pi^-]_{\rho^0}$; and
one pion misidentified as a kaon and one pion is not reconstructed, such as $D^0 [\pi^+\pi^+\pi^-]_{a_1^+}$.
\end{itemize}

The $\Delta E$ PDFs are parameterized by
a double Gaussian for the signal,
an exponential function for the $D$ true $B\bar{B}$ background,
an exponential function for the $D$ fake $B\bar{B}$ background,
a linear function for the $q\bar{q}$ background,
a double Gaussian for the $\bar{D}^0\rho^0$ background,
and a Gaussian for the $\bar{D}^0a_1^+$ background.
The $M_{\rm bc}$ PDFs are
a Gaussian for signal,
a Crystal Ball function for $D$ true $B\bar{B}$ background,
an Argus function for $D$  fake $B\bar{B}$ background,
an Argus function for $q\bar{q}$ background,
a sum of Gaussian and Argus function for $\bar{D}^0\rho^0$ background
and a Gaussian for $\bar{D}^0a_1^+$ background.
The $C'_{\rm NB}$ PDF are a sum of Gaussian and bifurcated Gaussian for each component.
The shape parameters of the  PDFs are fixed from MC samples.

\begin{figure}[]
  \begin{center}
  \includegraphics[width=1.0\textwidth]{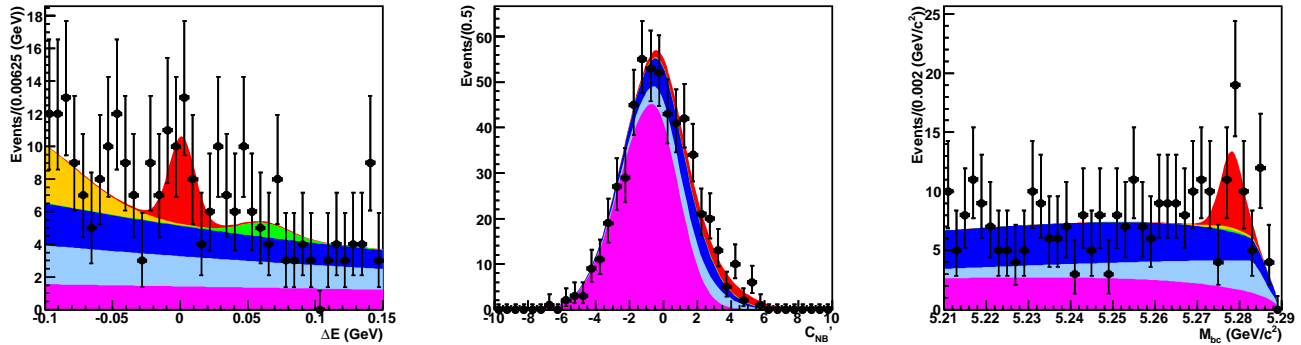}
    \caption{Projection of the fit to real data using the full Dalitz plot.
	Left: $\Delta E$ distribution with $M_{\rm bc} > 5.270~{\rm GeV}/c^{2}$ and $C'_{\rm NB} > 2$ requirements.
    	Middle: $C'_{\rm NB}$ distribution with $\mid\Delta E\mid < 0.030~{\rm GeV}$ and $M_{\rm bc} > 5.270~{\rm GeV}/c^{2}$ requirements.
    	Right: $M_{\rm bc}$ distribution with $\mid\Delta E\mid < 0.030~{\rm GeV}$ and $C'_{\rm NB} > 2$ requirements.
	Histograms show the fitted signal and background contributions,
	(red is signal,
	yellow is $D^0a_1^+$,
	green is $D^0\rho^0$,
	blue is $D$ fake $B\bar{B}$,
	light blue is $D$ true $B\bar{B}$ and
	magenta is $q\bar{q}$)
	and points with error bars are the data.
	}
    \label{fig:fit_data}
  \end{center}
\end{figure}

The Dalitz plot distributions of the background components are discussed in the next section.
The numbers of events in each bin can be free parameters in the fit,
thus there will be no uncertainty due to the modeling of the background distribution over the Dalitz plot in such an approach.
This procedure is justified for background that is either well separated from the signal
(such as peaking $B\bar{B}$ background),
or is constrained by a much larger number of events than the signal
(such as the $q\bar{q}$ background).
The results of the fit over the full Dalitz plot are shown in Figs.~\ref{fig:fit_data}.
We obtain a total of $44.2 ^{+13.3}_{-12.1}$ signal events.
The statistical significance is $2.8 \sigma$.

\section{DATA FIT IN BINS}

A combined fit is performed to obtain the $B^0\to DK^{*0}$ yield in each bin.
The combined fit constrains the amount of the $D$ true $B\bar{B}$ background in bins from the ratio of $D^0$ ($K_i$) and $\bar{D^0}$ ($K_{-i}$) from the generic MC, the amount of the $D$ fake $B\bar{B}$, $q\bar{q}$, $D^0\rho^0$ and $D^0a_1^+$ backgrounds in bins from the MC,
and takes the ($x_\pm$, $y_\pm$) variables as free parameters.
Fits to $B^0$ and $\bar{B}^0$ data are performed separately.
The plots illustrating the combined fit results are given in the Appendix.
In this fit, the additional free parameters are the total yields of $D$ true $B\bar{B}$, $D$ fake $B\bar{B}$, $q\bar{q}$ and peaking $B\bar{B}$ backgrounds for all over the Dalitz plane.
The values of ($x$, $y$) parameters and their statistical correlations obtained from the combined fit for signal sample are given in Table~\ref{tab:xy_data}.
\begin{table}
	\begin{center}
	\begin{tabular}{|c|c|}
	\hline
Parameter		& 					\\
\hline
$x_-$		& $+0.29 \pm 0.32$		\\
$y_-$		& $-0.33 \pm 0.41$		\\
corr.($x_-$, $y_-$)	& $+7.0 \%$		\\
\hline
$x_+$		& $+0.07 \pm 0.42$		\\
$y_+$		& $+0.05 \pm 0.45$		\\
corr.($x_+$, $y_+$)	& $-7.5 \%$		\\
\hline
	\end{tabular}
	\caption{($x$, $y$) parameters and their statistical correlations from combined fit of the $B^0\to DK^{*0}$ sample.
		The error is statistical uncertainty.
		The values and errors are obtained from likelihood distribution.}
	\label{tab:xy_data}
	\end{center}
\end{table}

\section{SYSTEMATIC UNCERTAINTY}

The systematic uncertainties on ($x$, $y$) are obtained by taking variations from the default procedure under differing assumptions.
Most systematic uncertainties are negligible compared to the statistical uncertainty. Therefore, numerical values are not given unless they are greater than or equal to 0.1 for any contribution.
There is an uncertainty due to the Dalitz plot efficiency variation
because of the difference in average efficiency over each bin for the flavor-tagged and $B^0\to DK^{*0}$ samples.
A maximum difference of 1.5 \% is obtained in a MC study.
The uncertainty is taken as the maximum of two quantities:
\begin{itemize}
\item the root mean square of $x$ and $y$ from smearing the numbers of events in the flavor-tagged sample $K_i$ by 1.5\%, or
\item the bias on $x$ and $y$ between the fits with and without efficiency correction for $K_i$ obtained from signal MC.
\end{itemize}
The uncertainty due to crossfeed of events between bins is estimated by taking the bias between the fits with and without
the correction.
The uncertainties due to the fixed parametrization of the signal and background PDFs are estimated by varying them by $\pm 1 \sigma$.
The uncertainty due to the $C'_{\rm NB}$ PDF distributions for ${B\bar{B}}$ are estimated by replacing them with the signal $C'_{\rm NB}$ PDF.
The uncertainty due to $D$ true and $D$ fake $B\bar{B}$ fraction of yield is estimated by varying then from 0 to 1.
The uncertainty due to PDFs shape is $(\Delta x_-, \Delta y_-, \Delta x_+, \Delta y_+) = (^{+0.0}_{-0.1}, ^{+0.1}_{-0.0}, ^{+0.0}_{-0.1}, ^{+0.0}_{-0.1})$.
The uncertainty arising from the finite sample of flavor-tagged $D\to K_S^0\pi\pi$ decays is evaluated by varying the value of $K_i$ within their statistical uncertainties.
The uncertainty due to the uncertainty on $k$ in Eq.~(\ref{eq:rS_delS_k}) is evaluated by varying the value of $k$ within its error~\cite{BaBar_k}.
The uncertainty due to the limited precision of $c_i$ and $s_i$ parameters is obtained by smearing the $c_i$ and $s_i$ values within their total errors and repeating the fits for the same experimental data.
The uncertainty due to $c_i$, $s_i$ is $(\Delta x_-, \Delta y_-, \Delta x_+, \Delta y_+) = (\pm0.0, \pm0.1, \pm0.1, \pm0.1)$.

The total systematic uncertainty is $(\Delta x_-, \Delta y_-, \Delta x_+, \Delta y_+) = (^{+0.0}_{-0.1}, \pm0.1, \pm0.1, \pm0.1)$.
The systematic uncertainty without $c_i$, $s_i$ is $(^{+0.0}_{-0.1}, ^{+0.1}_{-0.0}, ^{+0.0}_{-0.1}, ^{+0.0}_{-0.1})$.

\section{RESULT FOR ($x$, $y$) and $r_S$}

We use the frequentist approach with the Feldman-Cousins ordering~\cite{FC}
to obtain the physical parameters $\mu = (\phi_3, r_S, \delta_S)$ (or true parameters $\mu = z_{\rm true} = (x_-, y_-, x_+, y_+$))
from the measured parameters $z = z_{\rm meas} = (x_-, y_-, x_+, y_+)$ taken from the likelihood distribution.
In essence, the confidence level $\alpha$ for a set of physical parameters $\mu$ is calculated as
\begin{eqnarray}
\alpha(\mu) = \frac{\int_{{\cal D}(\mu)}p(z\mid \mu)dz}{\int_{\infty}p(z\mid \mu)dz},
	\label{eq:CL}
\end{eqnarray}
where $p(z\mid \mu)$ is the probability density to obtain the measurement result $z$ given the set of true parameters $\mu$.
The integration domain ${\cal D}(\mu)$ is given by the likelihood ratio (Feldman-Cousins) ordering:
\begin{eqnarray}
\frac{p(z\mid\mu)}{p(z\mid\mu_{\rm best}(z))} > \frac{p(z_0\mid\mu)}{p(z_0\mid\mu_{\rm best}(z_0))},
	\label{eq:int_reg}
\end{eqnarray}
where $\mu_{\rm best}(z)$ is $\mu$ that maximizes $p(z\mid\mu)$ for the given $z$, and $z_0$ is the result of the data fit.
This PDF is taken from MC pseudo-experiments.
 
Systematic errors in $\mu$ are obtained by varying the measured parameters $z$ within their systematic errors assuming a Gaussian distribution.
In this calculation we assume that the systematic errors are uncorrelated between the $B^0$ and $\bar{B}^0$ samples.

\begin{figure}[]
  \begin{center}
\includegraphics[width=0.41\textwidth]{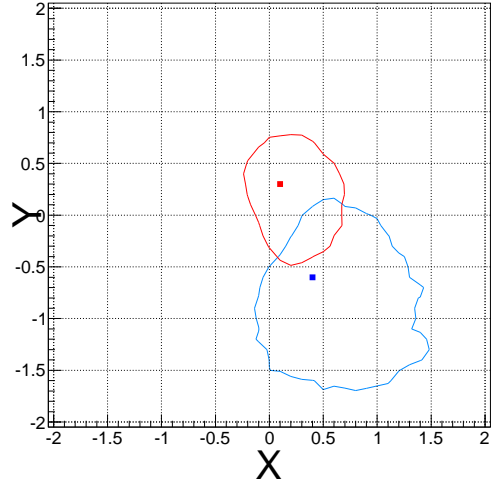}
    \caption{C.L. on ($x_-$, $y_-$) (blue) and ($x_+$, $y_+$) (red).
    	Dots show the most probable ($x$, $y$) values,
	lines show 68 \% contours.
	The fluctuations come from statistics of pseudo-experiments and C.L. step used.
    	}
    \label{fig:xy_CL}
  \end{center}
\end{figure}

\begin{figure}[]
  \begin{center}
    \includegraphics[width=1.\textwidth]{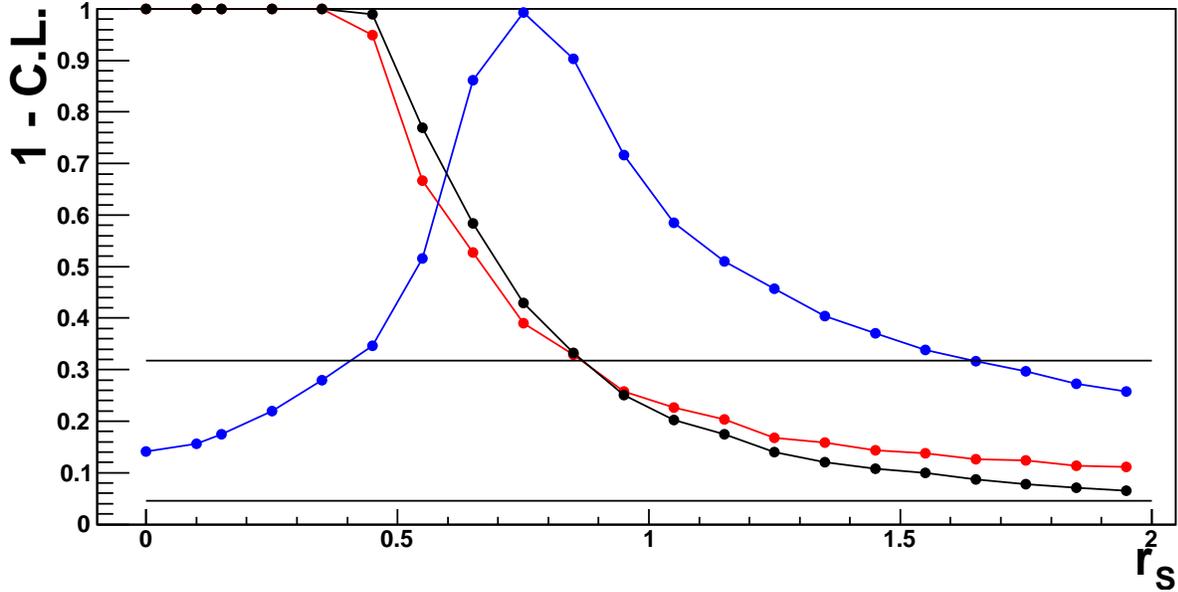}
    \caption{Likelihood profile for $r_S$.
	Blue is for  $\bar{B}^0$ ($x_-$, $y_-$),
	red is for $B^0$ ($x_+$, $y_+$)
	and black is $\bar{B}^{0}$ and $B^{0}$ combined.
    	}
    \label{fig:rS_CL}
  \end{center}
\end{figure}

As a result of this procedure, we obtain the confidence levels (C.L.) for ($x$, $y$) and the physical parameter $r_S$.
The C.L. contours on ($x$, $y$) are shown in Fig.~\ref{fig:xy_CL}.
The likelihood profile for $r_S$ is shown in Fig.~\ref{fig:rS_CL}.
The final results are:
\begin{eqnarray}
x_- &=& +0.4 ^{+1.0 +0.0}_{-0.6 -0.1} \pm0.0	\\
y_- &=& -0.6 ^{+0.8 +0.1}_{-1.0 -0.0} \pm0.1	\\
x_+ &=& +0.1 ^{+0.7 +0.0}_{-0.4 -0.1} \pm0.1	\\
y_+ &=& +0.3 ^{+0.5 +0.0}_{-0.8 -0.1} \pm0.1 ,	\\
r_S &<& 0.87 \hspace{30pt} {\rm at}~68 \% ~{\rm C.L.}.
\end{eqnarray}

\section{CONCLUSION}

We report the result of a measurement of the amplitude ratio $r_S$
using a model-independent Dalitz plot analysis of $D\to K_S^0\pi^+\pi^-$ decay
in the process $B^0\to DK^{*0}$. The value of $r_S$ indicates the sensitivity of the decay to $\phi_3$ because the statistical uncertainty is proportional to $1/r_S$. 
The measurement was performed with the full data sample of $711~{\rm fb}^{-1}$ ($772 \times 10^6 ~ B\bar{B}$ pairs)
collected by the Belle detector at the $\Upsilon(4S)$ resonance.
Model independence is achieved by binning the Dalitz plot of the $D\to K_S^0\pi^+\pi^-$ decay
and using the strong-phase coefficients for bins measured by the CLEO experiment~\cite{cisi_CLEO_2}.
We obtain the value $r_S < 0.87$ at 68 \% C.L.

This analysis is the first using the model-independent Dalitz technique on neutral $B$ mesons.
This measurement has resulted in lower statistical precision than the model-dependent measurement from BaBar with the $B^0\to DK^{*0}$ mode~\cite{Dalitz_BaBar_result}
despite the larger data sample due to the smaller $B^0\to DK^{*0}$ signal observed.
The result is consistent with the most precise $r_S$ measurement reported by the LHCb Collaboration~\cite{r_S_LHCb}  of $r_S = 0.240^{+0.055}_{-0.048}$ which uses $B^0\to [K^+K^-, K^\pm \pi^\mp, \pi^+\pi^-]_D K^{*0}$ decays.

\newpage

\appendix{APPENDIX}

\begin{figure}[]
  \begin{center}
 \includegraphics[width=1.0\textwidth]{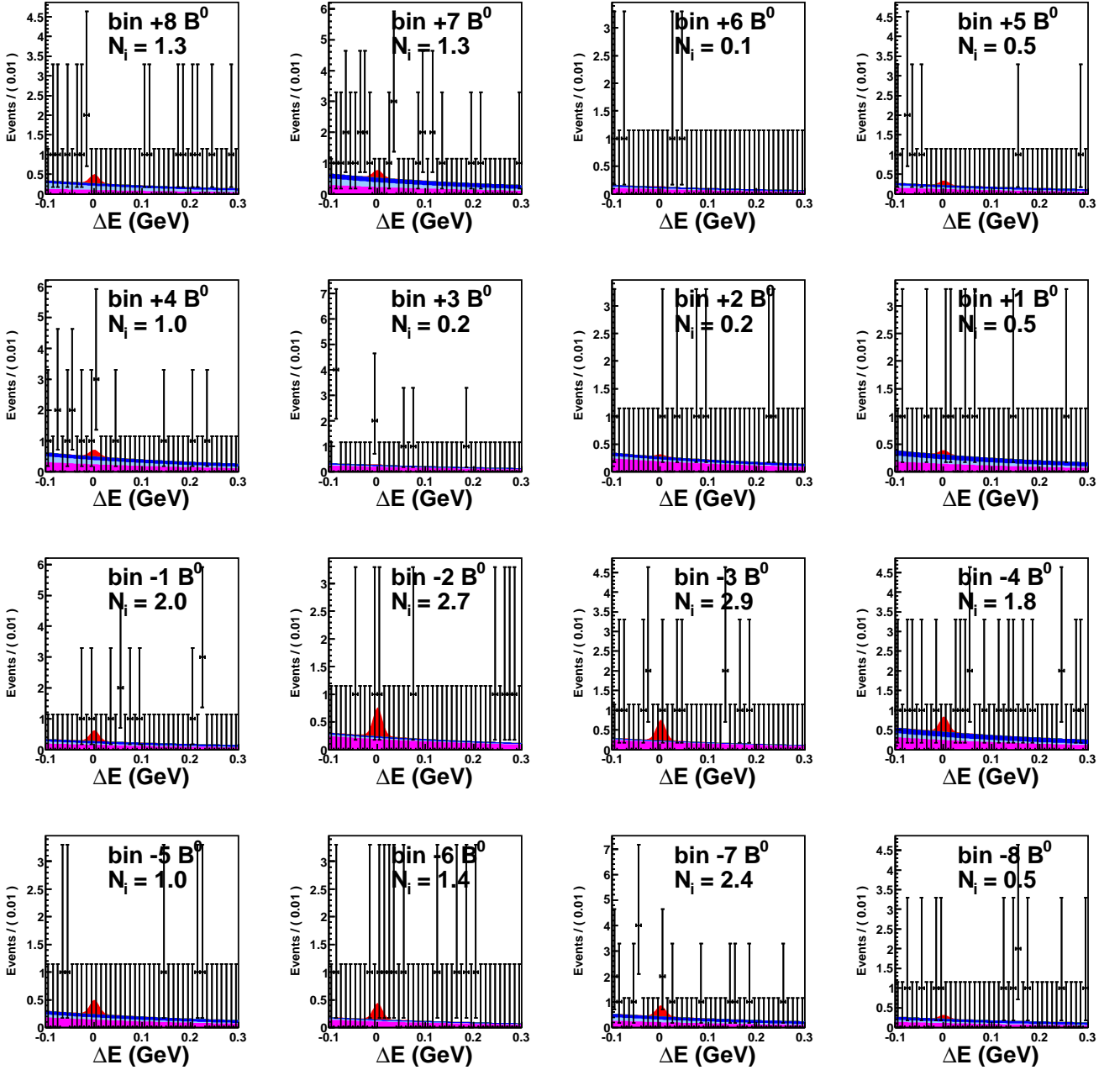}
    \caption{Projections of the combined fit of the $B^0\to DK^{*0}$ sample on $\Delta E$ for each Dailtz plot bin,
    	with the $M_{\rm bc} > 5270~{\rm MeV/c^2}$ and $C'_{\rm NB} > 2$. requirements.
	The fill styles for the signal and background components are the same as in Fig.~\ref{fig:fit_data}.
	}
    \label{fig:B0_binned}
  \end{center}
\end{figure}

\begin{figure}[]
  \begin{center}
 \includegraphics[width=1.0\textwidth]{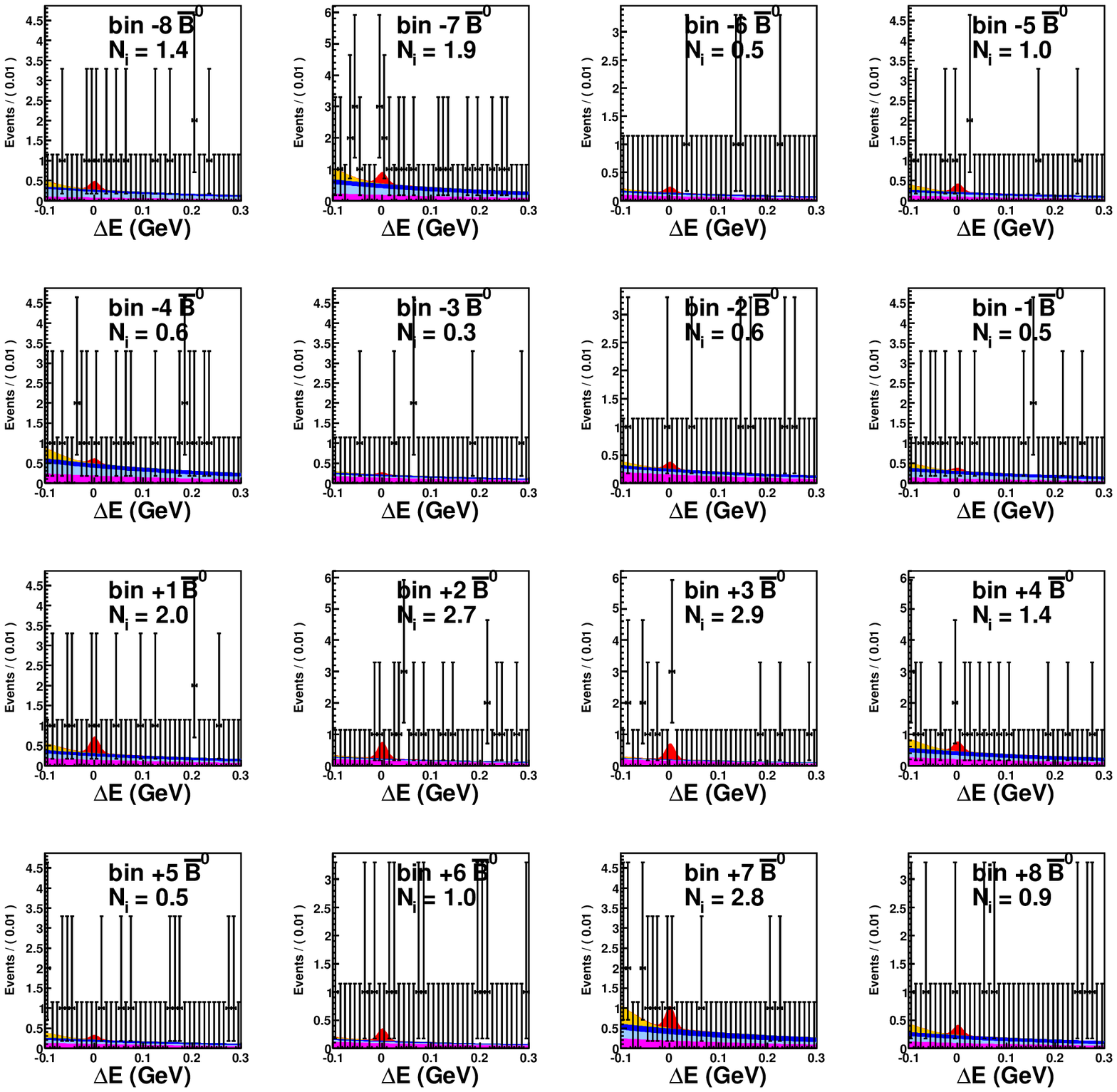}
    \caption{Projections of the combined fit of the $\bar{B}^0\to D\bar{K}^{*0}$ sample on $\Delta E$ for each Dailtz plot bin,
    	with the $M_{\rm bc} > 5270~{\rm MeV/c^2}$ and $C'_{\rm NB} > 2$. requirements.
	The fill styles for the signal and background components are the same as in Fig.~\ref{fig:fit_data}.
	}
    \label{fig:B0b_binned}
  \end{center}
\end{figure}

The results of the combined fit to $B^0\to DK^{*0}$ and $\bar{B}^0\to DK^{*0}$ samples separately
for each bin of the Dalitz plot are shown in Figs.~\ref{fig:B0_binned} and \ref{fig:B0b_binned}, respectively.
The plots show the projections of the data and the fitting model on the $\Delta E$ variable,
with the additional requirements $M_{\rm bc} > 5270~{\rm MeV/c^2}$ and $C'_{\rm NB} > 2$.

\end{document}